\documentclass[nofootinbib,preprintnumbers,prd,superscriptaddress,twocolumn,showpacs]{revtex4-1}

\usepackage{amsmath}
\usepackage{amsfonts}
\usepackage{amssymb}
\usepackage{graphicx}
\usepackage[titletoc]{appendix}
\usepackage{color}
\usepackage{hyperref}
\usepackage{cleveref}
\usepackage[rightcaption]{sidecap}
\usepackage{subfigure}
\usepackage{floatrow}
\usepackage{comment}
\usepackage{soul}
\usepackage{cancel}
\usepackage{dcolumn}
\usepackage{enumerate}
\usepackage[shortlabels]{enumitem}

\usepackage{array}
\usepackage{ctable}
\usepackage{multirow}
\usepackage{siunitx}
\usepackage{longtable}
\usepackage{tabularx}
\usepackage{booktabs}
\usepackage{float}

\graphicspath{{Graphics/}}

\def\be{\begin{equation}}
\def\ee{\end{equation}}
\def\bea{\begin{eqnarray}}
\def\eea{\end{eqnarray}}

\def\prl{Phys. Rev. Lett.}

\def\prd{Phys. Rev. D}
\def\mnras{MNRAS}
\def\aj{AJ}
\def\apj{ApJ}
\def\apjl{ApJ Lett.}
\def\apjs{ApJ Suppl. Ser.}

\def\aap{A\&A}

\def\physrep{Phys. Rep.}
\def\jcap{JCAP}

\def\apss{Astrophysics and Space Science}

\definecolor{vividviolet}{rgb}{0.62, 0.0, 1.0}
\definecolor{amaranth}{rgb}{0.9, 0.17, 0.31}
\definecolor{palatinateblue}{rgb}{0.15, 0.23, 0.89}
\definecolor{brightpink}{rgb}{1.0, 0.0, 0.5}
\definecolor{cornflowerblue}{rgb}{0.39, 0.58, 0.93}
\definecolor{deepcarminepink}{rgb}{0.94, 0.19, 0.22}
\definecolor{radicalred}{rgb}{1.0, 0.21, 0.37}

\hypersetup{ linktoc=all,
    colorlinks, linkcolor={palatinateblue},
    citecolor={brightpink}, urlcolor={amaranth}
}

\DeclareUnicodeCharacter{2212}{\ensuremath{-}}

\begin{document}

\title{Dark energy constraints using gamma-ray burst correlations with DESI 2024 data}

\author{Anna Chiara Alfano}
\email{a.alfano@ssmeridionale.it}
\affiliation{Scuola Superiore Meridionale, Largo S. Marcellino 10, 80138 Napoli, Italy.}
\affiliation{Istituto Nazionale di Fisica Nucleare (INFN), Sezione di Napoli Complesso Universitario Monte S. Angelo, Via Cinthia 9 Edificio G, 80138 Napoli, Italy.}

\author{Orlando Luongo}
\email{orlando.luongo@unicam.it}
\affiliation{Universit\`a di Camerino, Divisione di Fisica, Via Madonna delle carceri 9, 62032 Camerino, Italy.}
\affiliation{SUNY Polytechnic Institute, 13502 Utica, New York, USA.}
\affiliation{INFN, Sezione di Perugia, Perugia, 06123, Italy.}
\affiliation{INAF - Osservatorio Astronomico di Brera, Milano, Italy.}
\affiliation{Al-Farabi Kazakh National University, Al-Farabi av. 71, 050040 Almaty, Kazakhstan.}

\author{Marco Muccino}
\email{marco.muccino@lnf.infn.it}
\affiliation{Universit\`a di Camerino, Divisione di Fisica, Via Madonna delle carceri 9, 62032 Camerino, Italy.}
\affiliation{Al-Farabi Kazakh National University, Al-Farabi av. 71, 050040 Almaty, Kazakhstan.}
\affiliation{ICRANet, Piazza della Repubblica 10, 65122 Pescara, Italy.}

\begin{abstract}
Even though the Dark Energy Spectroscopic Instrument (DESI) mission does not exclude a dynamical dark energy evolution, the concordance paradigm, i.e., the $\Lambda$CDM model, remains statistically favored, as it depends on the fewest number of free parameters. In this respect, high redshift astrophysical sources, such as gamma-ray bursts, represent a formidable tool to model the form of dark energy, since they may provide a link between early and local redshift regimes. Hence, the use of these objects as possible distance indicators turns out to be essential to investigate the cosmological puzzle. To this end, we adopt two gamma-ray burst linear correlations, namely the $L_p-E_p$ and $L_0-E_p-T$ relations, to test the flat and non-flat $\Lambda$CDM, $\omega_0$CDM, and $\omega_0\omega_1$CDM cosmological models, i.e., those directly examined by the DESI collaboration. The inferred correlation coefficients and cosmological parameters are thus obtained by considering two independent Monte Carlo Markov chain analyses, the first considering the whole DESI data set and the second excluding a seemingly problematic data point placed at $z_{eff} = 0.51$. Using model selection criteria, the two above correlations do not show a preference on a precise cosmological model although, when the data point at $z_{eff}$ is included, the concordance paradigm appears to be the least favored among the tested cosmological models.
\end{abstract}



\maketitle
\tableofcontents

\section{Introduction}

Since the discovery of cosmic acceleration \cite{1998AJ....116.1009R, 1999ApJ...517..565P}, several dark energy models have been proposed, generalizing the assumption of a pure cosmological constant $\Lambda$ \cite{carroll1992cosmological,2001LRR.....4....1C,2003RvMP...75..559P,2001IJMPD..10..213C,2003PhRvL..90i1301L, 2005PhRvD..72f3501U, 2006IJMPD..15.1753C,2022PDU....3601045C, 2017PhR...692....1N, 2011PhR...505...59N,2024arXiv240917019W,2024PhRvD.110h3528W,2023PhRvD.108j3519W,2012Ap&SS.342..155B}.

Remarkably, DESI findings seem to indicate a preference for a dynamical dark energy contribution rather than a  cosmological constant, $\Lambda$ \cite{2024arXiv240403002D}. To back up DESI's claim, it is essential to investigate further in the behavior of dark energy by incorporating additional distance indicators, expanding our reach far beyond Type Ia Supernovae (SNe Ia) into higher redshift domains\footnote{The $\Lambda$CDM model, with the fewest number of free parameters, appears statistically favored and  works fairly well in characterizing the late-time dynamics, despite the fine-tuning and coincidence problems \cite{2006IJMPD..15.1753C} and cosmic tensions on the Hubble constant $H_0$ and the amplitude of clustering $S_8$ \cite{2022JHEAp..34...49A, 2021APh...13102605D, 2021APh...13102604D}.
Possibly, the $\Lambda$CDM can be theoretically incomplete, but mathematically well-behaved, as pointed out, e.g., in Refs. \cite{2018PhRvD..98j3520L,2024PDU....4401458B,2023CQGra..40j5004B,2016PhRvD..94h3525D}, exhibiting a transition time compatible with model-independent bounds \cite{2022MNRAS.509.5399C}.
}.

Thus, additional astrophysical sources covering a wider redshift range, \textit{i.e.}, up to $z\simeq 9$, turn out to be essential and, in this respect, gamma-ray bursts (GRBs) \cite{2011ApJ...736....7C, 2012ApJ...749...68S} represent high-energetic probes that can be used as distance indicators, as long as one circumnavigates the detrimental {\it circularity problem} \cite{2021Galax...9...77L}.
To this end, model-independent techniques can be worked out to elude the problem, enabling GRBs to be used as cosmological indicators \cite{2019MNRAS.486L..46A, 2023MNRAS.518.2247L, 2024arXiv240802536A, 2024JHEAp..42..178A, 2020A&A...641A.174L, 2021MNRAS.503.4581L}.

Since GRBs are found at intermediate redshifts, their study is essential in order to connect early- and late-time epochs and investigate further the nature of dark energy. Albeit typically providing larger values of mass density, encouraging results have been recently found based on the use of GRB correlations \cite{2021JCAP...09..042K, 2024arXiv240519953C}.
Indeed, several correlations have been proposed to connect different GRB observables and, conventionally, they are classified into
\begin{itemize}
    \item [-] prompt emission correlations \cite{2002A&A...390...81A, 2004ApJ...613L..13G, 2004ApJ...609..935Y};
    \item [-] prompt-afterglow emission correlations \cite{2005ApJ...633..611L, 2008MNRAS.391L..79D, 2011MNRAS.418.2202D, 2019ApJS..245....1T, 2012MNRAS.425.1199B, 2015A&A...582A.115I}.
\end{itemize}

Driven by the aim of intermediate redshift investigations, we resort the widely used \textit{B\'ezier parametric interpolation} to render GRBs model-independent \cite{2019MNRAS.486L..46A}. Hence, we here adopt this technique to calibrate the prompt emission $L_p-E_p$ (or {\it Yonetoku}) correlation \cite{2004ApJ...609..935Y} and the prompt-afterglow $L_0-E_p-T$ (or {\it Combo}) correlation \cite{2015A&A...582A.115I} using observational Hubble data (OHD) and the baryonic acoustic oscillations (BAO) sample inferred by DESI. The reasons behind choosing these correlation functions are multiple, say:
\begin{itemize}
    \item[-] we want to compare a prompt emission correlation with a prompt-afterglow emission one to check if model selection criteria prefer the same outcome,
    \item[-] we intend to work alternative correlations, based on different  calibrated data sets,
    \item[-] we seek deviations from the usual $E_p-E_{iso}$ correlation, investigated in Ref. \cite{2024arXiv240802536A},
    \item[-] we aim at checking the DESI findings since recent analyses involving GRB samples seem to go into the direction of DESI claims when they are combined with other probes such as SNe Ia or BAO\footnote{In Ref. \cite{2020A&A...641A.174L} the authors found that a cosmographic study of the $L_p-E_p$ and $L_0-E_p-T$ correlations calibrated via OHD and in conjuction with SNe Ia and BAO showed to favor a midly evolving dark energy against the concordance paradigm. Additionally, in Ref. \cite{2021MNRAS.503.4581L} using a machine learning-driven approach, the calibrated $L_0-E_p-T$ correlation, together with SNe Ia and BAO, also seems to indicate a non-constant dark energy behavior.}.
\end{itemize}

Hence,  the procedure adopted in this work to calibrate the correlations consists in approximate both the Hubble rate and the luminosity distance to have a totally model-free analysis. In addition, we perform two independent fits where in the first we consider all the data points from DESI while in the second we exclude a seemingly problematic data point at $z_{eff} = 0.51$\footnote{The data point is inferred through the luminous red galaxy (LRG) sample. The discussed data point falls in the redshift range $0.4<z<0.6$ and it is usually labeled as LRG1 \cite{2024arXiv240403002D}.} criticized for a drastical increase of the $\Omega_m$ value, as pointed out by Ref. \cite{2024arXiv240408633C}\footnote{This trend is also seen in the Dark Energy Survey five-year SNe data (DES 5YR SNe) considering a tomographic approach \cite{2024arXiv240606389C}.}. Accordingly, we test different cosmological models combining calibration and cosmological GRB samples to find the best-fit coefficients for each correlation and the best-fit cosmological parameters for each model, respectively. Motivated by the DESI findings, we focus on the three most sensitive cosmological models, analyzed by DESI, namely
\begin{itemize}
    \item [-] the concordance  $\Lambda$CDM paradigm, where the dark energy fluid exactly corresponds to $\Lambda$ \cite{2021A&A...652C...4P};
    \item [-] the $\omega_0$CDM model in which the equation of state correponds to a constant barotropic factor $\omega_0$  \cite{2006IJMPD..15.1753C};
    \item [-] the $\omega_0\omega_1$CDM model where the barotropic factor evolves with $z$, {\it i.e.}, $\omega(z) = \omega_0+\omega_1 z(1+z)^{-1}$ \cite{2001IJMPD..10..213C, 2003PhRvL..90i1301L}.
\end{itemize}

For all the above cases, we investigate spatially flat and non-flat scenarios.
We select the most viable cosmological frameworks working out Monte Carlo -- Markov chain (MCMC) analyses, based on the Metropolis-Hastings algorithm \cite{1953JChPh..21.1087M, 1970Bimka..57...97H}, using $L_p-E_p$ and $L_0-E_p-T$ correlations.
We perform two independent fits for both correlations. In the first we consider the whole DESI data set, while in the second option we exclude the contested data point placed at $z_{eff} = 0.51$ \cite{2024arXiv240408633C}. In both the cases, we bound the correlation coefficients and the cosmological parameters for the above three models, with $\Omega_k=0$ and $\Omega_k\neq0$. Afterwards, we compare our findings with those obtained by the Planck collaboration \cite{2021A&A...652C...4P} and the Sloan Digital Sky Survey (SDSS) \cite{2021PhRvD.103h3533A}. Last but not least, for both correlations, we statistically investigate for the most favored model through the use of selection criteria. As it will be clear throughout the text, it turns out that none of the correlations here-adopted favors a particular dark energy model, but cannot exclude a slightly evolving scenario, leaving open the possibility that dark energy evolves in time, even at late-times.

The work is structured as follows. In Sect. \ref{sec1} we introduce the GRB correlations involved in our analysis. We also describe the OHD and DESI-BAO data sets used in Sect. \ref{sec3} to calibrate the correlations of our choosing to make the analysis fully model-independent. In Sect. \ref{sec4} we present the numerical outcomes of our MCMC computation. Finally, in Sect. \ref{sec5} we draw our conclusions and propose further developments.

\section{GRB correlations and cosmic data sets}\label{sec1}

All GRB correlations follow a linear behaviour in the $\log$--$\log$ plane connecting different observables. These correlations are mainly divided into two classes:

\begin{itemize}
    \item [-] {\bf Prompt emission correlations}, that combine prompt emission quantities, typically observed in the hard-X/$\gamma$-ray energy domains, within the burst duration defined by the $t_{90}$ time \cite{2021Galax...9...77L}. The burst spectral energy distribution, integrated over $t_{90}$, is generally best-fit by the so-called Band model \cite{1993AIPC..280..872B}.
    \item [-] {\bf Prompt-afterglow emission correlations}, that combine prompt and afterglow emission observables. The afterglow emission, observed in the X-rays after $t\sim100$ s from the on-set of the prompt emission, is characterized by a typical light curve exhibiting an early steep decay (coinciding with the end of the prompt emission), followed by a plateau phase and a later, more gentle final decay \cite{2021Galax...9...77L}.
\end{itemize}

In the following, we introduce the prompt emission $L_p$--$E_p$ and the prompt-afterglow emission $L_0$--$E_p$--$T$ correlations and summarize their main associated features.

\subsection{The $L_p-E_p$ correlation}

The $L_p-E_p$ or \textit{Yonetoku} \cite{2004ApJ...609..935Y} falls in the above class of prompt emission correlations \cite{2002A&A...390...81A, 2004ApJ...613L..13G}.
This linear correlation, with slope $a$ and intercept $b$, is expressed by
\begin{equation}
    \log\left(\frac{L_p}{\text{erg}\cdot\text{s}^{-1}}\right) = 52 + b + a\log\left(\frac{E_p}{\text{keV}}\right),
\end{equation}
connects the rest-frame spectral peak energy $E_p = E_p^0(1+z)$, where $E^0_p$ is the observed spectral peak energy got from the Band model, and the peak luminosity $L_p$ got from the brightest $1$ s interval of the GRB prompt light curve interval. The luminosity peak is defined as
\begin{equation}
\label{lpeak}
    L_p = 4\pi D_L^2(z, p)F_p,
\end{equation}
where $F_p$ is the measured peak flux and $D_L(z, p)$ is the model-dependent luminosity distance -- which depends on the redshift $z$ and on the cosmological parameters $p$ -- usually jeopardized by the \textit{circularity problem}.

\subsection{The $L_0-E_p-T$ correlation}

The $L_0-E_p-T$ correlation \cite{2015A&A...582A.115I} belongs to the prompt-afterglow emission correlations  \cite{2005ApJ...633..611L, 2008MNRAS.391L..79D, 2011MNRAS.418.2202D, 2019ApJS..245....1T, 2012MNRAS.425.1199B}.
This correlation is obtained by combining the prompt emission $E_p-E_{iso}$ correlation and the prompt-afterglow emission $E^X_{iso}-E_{iso}-E_p$ correlation \cite{2002A&A...390...81A,2012MNRAS.425.1199B} and for this reason is referred as to as the \textit{Combo} correlation.

The Combo correlation, widely used in the literature with promising results in both cosmology \cite{muccino2020confront, 2021ApJ...908..181M, 2021MNRAS.503.4581L, 2022LRR....25....6M} and cosmography \cite{2020A&A...641A.174L, 2024JHEAp..42..178A} applications, is a linear correlationwith slope $a$ and intercept $b$, that reads
\begin{equation}\label{combo}
    \log\left(\frac{L_0}{\text{erg}\cdot\text{s}^{-1}}\right) = b+a\log\left(\frac{E_p}{\text{keV}}\right)-\log\left(\frac{T}{\text{s}}\right),
\end{equation}
in which we defined $T\equiv \tau/|1+\alpha|$ and employed the luminosity $L_0$ and the rest-frame duration $\tau$ of the X-ray plateau, and the late decay index $\alpha$ of the afterglow.

Like all GRB correlations, Eq. \eqref{combo} is affected by the \textit{circularity problem} through the definition of $L_0$, i.e.,
\begin{equation}\label{lum}
    L_0 = 4\pi D_L^2(z, p)F_0,
\end{equation}
involving the X-ray flux $F_0$ of the plateau and $D_L(z, p)$.

\subsection{Calibration catalogs}\label{sec2}

To use GRB correlations as distance indicators requires a calibration procedure.
To do so, we resort two data sets
\begin{itemize}
    \item [-] the $N_O=34$ OHD data points \cite{2014RAA....14.1221Z, 2002ApJ...573...37J, 2005PhRvD..71l3001S, 2012JCAP...08..006M, 2016JCAP...05..014M, 2017MNRAS.467.3239R, 2010JCAP...02..008S, 2022ApJ...928L...4B, 2023ApJS..265...48J, 2023A&A...679A..96T, 2015MNRAS.450L..16M};
    \item [-] the $N_B=12$ BAO data points from the DESI \cite{2024arXiv240403002D}.
\end{itemize}

The OHD data points are essentially measurements of the Hubble rate at various redshifts $z$
\begin{equation}
    H(z) = -\frac{1}{1+z}\left(\frac{d t}{d z}\right)^{-1},
\end{equation}
obtained by relying on model-independent measurements of the ratio $dz/dt\simeq\Delta z/\Delta t$, namely
measuring $\Delta z$, through spectroscopic observations, and $\Delta t$, by assuming that in old passively evolving galaxies found at different redshift $z$ all the stars formed simultaneously, making them \textit{cosmic chronometers} \cite{2002ApJ...573...37J, 2018ApJ...868...84M}.

Finally, we consider the $12$ BAO observations from DESI. These data are divided into $7$ redshift bins and consider $6$ tracers encompassing bright galaxy survey (BGS), emission line galaxies (ELG), quasars (QSO), Lyman-$\alpha$ forest quasars (Ly$\alpha$ QSO), LRG and the combined LRG+ELG. In particular, LRG is divided into LRG1, LRG2 and LRG3, to distinguish the redshift intervals they are in \cite{2024arXiv240403002D}.

Using these tracers the values of the transverse comoving distance $D_M/r_d$, the Hubble rate distance $D_H/r_d$ and the angle-averaged distance $D_V/r_d$ are derived
\begin{subequations}
\label{DESIdata}
  \begin{align}
      &\frac{D_M(z)}{r_d} = \frac{1}{1+z}\frac{D_L(z)}{r_d},&\\
      &\frac{D_H(z)}{r_d} = \frac{1}{r_d}\frac{c}{H(z)},&\\
      &\frac{D_V(z)}{r_d} = \frac{1}{r_d}\left[\frac{cz}{H(z)}\right]^{1/3}\left[\frac{D_L(z)}{1+z}\right]^{2/3},&
  \end{align}
\end{subequations}
where $r_d$ is the sound horizon at the baryon drag epoch.

\section{Model-independent calibration of GRB correlations}\label{sec3}

In our approach, correlations can be calibrated through OHD and DESI-BAO catalogs resorting to the well-established \textit{B\'ezier} interpolation \cite{2019MNRAS.486L..46A}.

By B\'ezier interpolating the OHD catalog, we obtain the parametrization of the Hubble rate \cite{2023MNRAS.518.2247L, 2024arXiv240802536A, 2024JHEAp..42..178A, 2024A&A...686A..30A, 2023MNRAS.523.4938M}
\begin{subequations}
    \begin{align}\label{bezier}
 &H_n(x) = \sum^{n}_{i = 0} \text{100}\alpha_i h^i_n(x)\
\text{km}\cdot\text{s}^{-1}\cdot\text{Mpc}^{-1}, \\
&h^i_n(x) \equiv n!\frac{x^i}{i!}\frac{(1-x)^{n-i}}{(n-i)!},
\end{align}
\end{subequations}
where $\alpha_i$ are the \textit{B\'ezier coefficients} for the OHD catalog, the first coefficient $\alpha_0$ can be identified with the reduced Hubble rate $h_0$, \textit{i.e.}, $\alpha_0\equiv h_0$ and $0\leq x\equiv z/z_O \leq 1$ where $z_O = 1.965$ is the maximum redshift of the OHD catalog. Further, we stop at the order $n=2$ because at higher orders the interpolation does not behave anymore as a monotonically growing, non-linear function \cite{2021MNRAS.503.4581L}.

The B\'ezier interpolation of the luminosity distance of BAO data is achieved through
\begin{subequations}
\begin{align}\label{bezier1}
    &D^2_m(y) = \sum^{m}_{j=0}\text{10}^6\beta_jd^j_m(y)\ \text{Mpc}^2, \\
    &d^j_m(y)\equiv m!\frac{y!}{j!}\frac{(1-y)^{m-j}}{(m-j)!},
\end{align}
\end{subequations}
where $\beta_j$ are the \textit{B\'ezier coefficients} for the BAO catalog and $0\leq y\equiv z/z_B \leq 1$ where $z_B = 2.33$ is the maximum redshift of the BAO catalog. We stop at the interpolation order $m=3$ so that Eq. \eqref{bezier1} behaves as a monotically growing, non-linear function. Further, when $j=0$ we set $D^2_m(0)\equiv 0$ from the definition of distance \cite{2023MNRAS.518.2247L}.

To derive the \textit{B\'ezier coefficients} in Eqs. \eqref{bezier} -- \eqref{bezier1} we perform a MCMC simulation using the Metropolis-Hasting algorithm \cite{1953JChPh..21.1087M, 1970Bimka..57...97H} by maximizing the following log-likelihood
\begin{equation}
    \ln\mathcal{L} = \ln\mathcal{L}_O+\ln\mathcal{L}_B,
\end{equation}
where the subscript $O$ and $B$ are associated with OHD and BAO samples, respectively.

The OHD log-likelihood takes the following form
\begin{subequations}
    \begin{align}
    \ln\mathcal{L}_O &= -\frac{1}{2} \left[\chi^2_O + \sum^{N_O}_{i = 1} \ln\left(2\pi\sigma^2_{H_i}\right)\right], \\
    \chi^2_O &= \sum^{N_O}_{i = 1} \left[\frac{H_i-H_2(z_i)}{\sigma^2_{H_i}}\right]^2,
    \end{align}
\end{subequations}
where $H_i$ are the observed data with associated errors $\sigma_{H_i}$ and $H_2(z_i)$ is the B\'ezier-interpolated Hubble rate.

Conversely, the BAO log-likelihood takes the form
\begin{subequations}
    \begin{align}
    \ln\mathcal{L}_B &= -\frac{1}{2}\left[\chi^2_B + \sum^{N_B}_{i = 1} \ln\left(2\pi\sigma^2_{D_i}\right)\right], \\
    \chi^2_B &= \sum^{N_B}_{i = 1} \left[\frac{D_i-D(z_i)}{\sigma^2_{D_i}}\right]^2,
    \end{align}
\end{subequations}
where $D_i$ are the BAO data with errors $\sigma_{D_i}$ and $D(z_i)$ are the \textit{B\'ezier}-interpolated ratios defined in Eqs. \eqref{DESIdata}.

For our study, having already performed this computation in Ref. \cite{2024arXiv240802536A} we simply describe how the simulation was made and then we report the best-fit parameters of the \textit{B\'ezier} coefficients. Specifically, we performed two different kind of analyses, letting $r_d$ vary in the interval $r_d\in[138, 156]$ Mpc at steps of $2$ Mpc and considering
\begin{itemize}
    \item [1)] all the data points from DESI;
    \item [2)] the DESI sample without LRG1 at $z_{eff} = 0.51$ due to an higher value of $\Omega_m$ reported in Ref. \cite{2024arXiv240408633C}.
\end{itemize}

The best-fit results, obtained by maximizing the log-likelihoods in the two cases, are reported in Tab. \ref{tab:bfBezier}.

\begin{table*}
\centering
\setlength{\tabcolsep}{0.25em}
\renewcommand{\arraystretch}{1.7}
\begin{tabular}{lccccccc}
\hline
\hline
LGR1 & $r_d$     & $\alpha_0\equiv h_0$ & $\alpha_1$ & $\alpha_2$ & $\beta_1$ & $\beta_2$ & $\beta_3$ \\ \hline
Yes & $152$ & $0.718^{+0.022(0.048)}_{-0.020(0.054)}$ & $0.976^{+0.016(0.033)}_{-0.021(0.036)}$ & $1.936^{+0.006(0.012)}_{-0.007(0.016)}$ & $-1.333^{+1.262(2.400)}_{-0.640(1.742)}$ & $51.103^{+3.108(8.844)}_{-6.449(11.87)}$ & $404.383^{+18.090(37.317)}_{-9.648(25.727)}$\\

No & $142$ & $0.646^{+0.023(0.054)}_{-0.036(0.068)}$ & $1.085^{+0.025(0.044)}_{-0.014(0.033)}$ & $2.088^{+0.009(0.017)}_{-0.007(0.015)}$ & $-1.586^{+0.575(1.538)}_{-1.025(1.804)}$ & $42.190^{+4.819(9.644)}_{-4.855(9.512)}$ & $360.954^{+13.295(29.057)}_{-11.321(28.320)}$
 \\
\hline
\hline
\end{tabular}
\caption{Best-fit \textit{B\'ezier} coefficients with $1$-$\sigma$($2$-$\sigma$) error bars considering all BAO datapoints from DESI (upper panel) and without the LRG1 point (lower panel). The results are taken from Tabs. VII-VIII in our previous work, Ref. \cite{2024arXiv240802536A}.}
\label{tab:bfBezier}
\end{table*}

\subsection{The calibration procedure}

We are now able to calibrate our correlations. For the $L_p-E_p$ and $L_0-E_p-T$ correlations we substitute inside Eqs. \eqref{lpeak} and \eqref{lum} the luminosity distance reconstructed with \textit{B\'ezier} polynomials leading to
\begin{subequations}
    \begin{align}
         L_p^{cal} = 4\pi D_{3}^2(z, p)F_p,\\
         L_0^{cal} = 4\pi D^2_{3}(z, p)F_0.
    \end{align}
\end{subequations}
thus, bypassing the \textit{circularity problem} and the bias based of an underlying cosmological model.

We can now simultaneously determine both the correlation coefficients and the cosmological parameters of the examined model by considering two samples within the same GRB catalog: one that considers only GRBs at redshift $z\leq z_B$, providing constraints on the coefficients of the calibrated correlation though the log-likelihood $\ln\mathcal{L}^{cal}$, and the other one
one which considers the whole catalog, providing the constraints on the cosmological parameters through the log-likelihood $\ln\mathcal{L}^{cos}$. In this way the total log-likelihood takes the following form
\begin{equation}
    \ln\mathcal{L} = \ln\mathcal{L}_{cal} + \ln\mathcal{L}_{cos},
\end{equation}
The two log-likelihoods take the following forms
\begin{subequations}
\begin{align}
\label{logcal}
    \ln\mathcal{L}_{cal} &= -\frac{1}{2}\sum^{N_{cal}}_{j=1}\left[\chi^2_{cal}+\ln\left(2\pi\sigma^2_{\mathcal{Y}_i}\right)\right],\\
    \label{logcos}
    \ln\mathcal{L}_{cos} &= -\frac{1}{2}\sum^{N_{cos}}_{j=1} \left[\chi^2_{cos}+\ln\left(2\pi\sigma^2_{\mu_j}\right)\right],
\end{align}
\end{subequations}
where, for each correlation, $N_{cal}$ is the number of GRBs at $z\leq z_B$, whereas $N_{cos}$ is the size of the GRB samples.

Expliciting all quantities inside Eqs. \eqref{logcal}-\eqref{logcos} for both correlations we have:
\begin{itemize}
    \item [-] for the $L_p-E_p$ correlation, we define
\begin{subequations}
    \begin{align}
        X_j & = a\log E_{p,j}+b\,,\\
        \chi^2_{cal} & = \sum^{N_{cal}}_{j=1} \left[\frac{\log L_{p,j}- X_j-52}{\sigma_{\mathcal{Y}_i}}\right]^2,\\
    \chi^2_{cos} &= \sum^{N_{cos}}_{j=1} \left[\frac{\mu_0+2.5\left(X_j-\log F_{p,j}\right)-\mu(z_j)}{\sigma_{\mu_j}}\right]^2,\label{chicos1}
    \end{align}
\end{subequations}
and the uncertainties are given by
\begin{subequations}
\begin{align}
\sigma^2_{\mathcal{Y}_j} &= \sigma^2_{\log L_{p,j}}+a^2\sigma^2_{\log E_{p,j}}+\sigma^2_{ext},\\
\sigma^2_{\mu_j} &= 2.5^2\left(\sigma^2_{\mathcal{Y}_j} - \sigma^2_{\log L_{p,j}}+\sigma^2_{\log F_{p,j}}\right),
\end{align}
\end{subequations}
where $\mu_0=32.55$ converts from Mpc to cm.

\item [-] for the $L_0-E_p-T$ correlation we have
\begin{subequations}
    \begin{align}
     X_j & = a\log E_{p,j} - \log T_j +b\,,\\
     \chi^2_{cal} & = \sum^{N_{cal}}_{j=1} \left[\frac{\log L_{0,j}-X_j}{\sigma_{\mathcal{Y}_j}}\right]^2,\\
     \chi^2_{cos} &= \sum^{N_{cos}}_{j=1} \left[\frac{\mu_0+2.5\left(X_j-\log F_{0,j}\right)-\mu(z_j)}{\sigma_{\mu_j}}\right]^2.\label{chicos2}
    \end{align}
\end{subequations}
and the uncertainties are given by
\begin{subequations}
\begin{align}
\sigma^2_{\mathcal{Y}_j} &= \sigma^2_{\log L_{0,j}}+a^2\sigma^2_{\log E_{p,j}}+\sigma^2_{\log T_j}+\sigma_{ext}^2,\\
\sigma^2_{\mu_j} &= 2.5^2\left(\sigma^2_{\mathcal{Y}_j} - \sigma^2_{\log L_{0,j}} + \sigma^2_{\log F_{0,j}}\right),
\end{align}
\end{subequations}
where $\mu_0=-97.45$ converts from Mpc to cm.
\end{itemize}

Both correlations are characterized by an additional intrinsic extra-scatter term $\sigma^2_{ext}$  \cite{2005physics..11182D}.
Further, in Eqs. \eqref{chicos1} and \eqref{chicos2}, we considered the theoretical distance moduli $\mu(z)=25+5\log\left[D_L(z)/{\text{Mpc}}\right]$ that incorporate the model luminosity distance $D_L(z)$
\begin{equation}
    D_L(z) = \frac{c(1+z)}{100\alpha_0\sqrt{|\Omega_k|}}S_k\left[\int^z_0 \frac{\sqrt{|\Omega_k|}}{E(z^\prime)} dz^\prime\right],
\end{equation}
where $E(z)=H(z)/H_0$ is the normalized Hubble rate
\begin{equation}
E(z) = \sqrt{\Omega_m(1+z)^3+\Omega_k(1+z)^2+\Omega_{DE}f(z)},
\end{equation}
$f(z)$ depends upon the cosmological model under test, and the \textit{B\'ezier} coefficient $\alpha_0$ coincides with the reduced Hubble constant $h_0$. The element $S_k(x)$ changes according to the sign of the curvature parameter $\Omega_k$, $S_k(x) = \sinh(x)$ for a positive $\Omega_k$, $S_k(x) = \sin(x)$ for a negative $\Omega_k$ and $S_k(x) = x$ for a null $\Omega_k$ \cite{1995ApJ...450...14G}.

\section{Numerical results}\label{sec4}

In this section we present and physically discuss the numerical outcomes of our study for three cosmological models, namely $\Lambda$CDM, $\omega_0$CDM and $\omega_0\omega_1$CDM models.

First, we discuss the results got using the $L_p-E_p$ correlation and then the ones obtained from the $L_0-E_p-T$ correlation. We dedicate our attention to the study of the cosmological parameters of all the models taken under consideration deduced from our MCMC computations. We confront them with the values inferred from the Planck Collaboration \cite{2021A&A...652C...4P} and from the SDSS \cite{2021PhRvD.103h3533A}. To distinguish the values of the matter density $\Omega_m$ and of the curvature parameter $\Omega_k$ of the two surveys we denote with a superscript $P$ the ones referring to Planck while with a superscript $S$ the ones referring to the SDSS. In all models we dinstinguish the flat scenario ($\Omega_k=0$) from the non-flat one ($\Omega_k\neq 0$). The contour plots are made using pyGTC, a free Python package \cite{2016JOSS....1...46B}.

\subsection{Results for the $L_p-E_p$ correlation}

The best-fit correlation and cosmological parameters got from the use of the $L_p-E_p$ correlation are listed in Tab. \ref{tab:DESIBAOYonetoku1} and the corresponding contour plots are portrayed in Figs. \ref{plot:yonetokufull}-\ref{plot:yonetokunolrg1}.
The results are differentiated between flat and non-flat geometries and whether the LRG1 data point is considered or not.

\begin{table*}
\centering
\footnotesize
\setlength{\tabcolsep}{0.25em}
\renewcommand{\arraystretch}{1.6}
\begin{tabular}{ccccccc}

\hline\hline
\multicolumn{7}{c}{$L_p-E_p$ correlation}\\
\hline\hline
$a$ & $b$ & $\sigma$ & $\Omega_m$ & $\Omega_k$ & $\omega_0$ & $\omega_1$\\
\hline
\multicolumn{7}{c}{With LRG1}\\
\hline
\multicolumn{7}{c}{$\Lambda$CDM} \\
\cline{1-7}
$1.488^{+0.091(0.147)}_{-0.096(0.152)}$ & $-3.427^{+0.227(0.362)}_{-0.237(0.380)}$ & $0.340^{+0.035(0.061)}_{-0.035(0.057)}$ & $0.453^{+0.176(0.313)}_{-0.162(0.235)}$ & $-$ & $-1$ & $0$ \\
$1.473^{+0.083(0.141)}_{-0.090(0.144)}$ & $-3.396^{+0.222(0.364)}_{-0.209(0.352)}$ & $0.335^{+0.035(0.063)}_{-0.037(0.054)}$ & $0.433^{+0.188(0.337)}_{-0.139(0.209)}$ & $-0.513^{+0.349(0.817)}_{-0.230(0.372)}$ & $-1$  & $0$\\
\hline
\multicolumn{7}{c}{$\omega_0$CDM} \\
\cline{1-7}
$1.477^{+0.077(0.131)}_{-0.090(0.147)}$ & $-3.337^{+0.187(0.334)}_{-0.227(0.368)}$ & $0.332^{+0.037(0.060)}_{-0.034(0.053)}$ & $0.498^{+0.143(0.248)}_{-0.112(0.166)}$ & $-$ & $-5.961^{+3.030(3.787)}_{\rm unc(\rm unc)}$ &
$-$ \\
$1.450^{+0.094(0.156)}_{-0.089(0.153)}$ & $-3.276^{+0.182(0.358)}_{-0.280(0.437)}$ & $0.325^{+0.042(0.064)}_{-0.030(0.048)}$ & $0.622^{+0.237(0.367)}_{-0.198(0.347)}$ & $-0.356^{+0.304(0.601)}_{-0.308(0.509)}$ & $-5.104^{+4.573(5.039)}_{\rm unc(\rm unc)}$ &
$-$ \\
\hline
\multicolumn{7}{c}{$\omega_0\omega_1$CDM}\\
\cline{1-7}
$1.463^{+0.099(0.159)}_{-0.068(0.137)}$ & $-3.368^{+0.204(0.375)}_{-0.231(0.375)}$ & $0.331^{+0.037(0.063)}_{-0.034(0.053)}$ & $0.519^{+0.137(0.242)}_{-0.125(0.202)}$ & $-$ & $-4.683^{+3.791(4.472)}_{\rm unc(\rm unc)}$ &
$-5.851^{+2.359(5.244)}_{\rm unc(\rm unc)}$ \\
$1.443^{+0.096(0.155)}_{-0.074(0.132)}$ & $-3.290^{+0.185(0.348)}_{-0.245(0.396)}$ & $0.333^{+0.034(0.059)}_{-0.037(0.057)}$ & $0.651^{+0.205(0.333)}_{-0.205(0.344)}$ & $-0.378^{+0.393(1.461)}_{-0.283(0.476)}$ & $-2.530^{+1.494(1.788)}_{-1.913(2.440)}$ &
$+1.457^{+5.366(5.990)}_{-2.941(3.472)}$ \\
\hline
\hline
\multicolumn{7}{c}{Without LRG1}\\
\hline
\multicolumn{7}{c}{$\Lambda$CDM} \\
\cline{1-7}
$1.478^{+0.082(0.149)}_{-0.095(0.146)}$ & $-3.458^{+0.235(0.362)}_{-0.202(0.375)}$ & $0.334^{+0.036(0.061)}_{-0.035(0.053)}$ & $0.753^{+0.300(0.574)}_{-0.250(0.344)}$ & $-$ & $-1$ & $0$ \\
$1.445^{+0.105(0.167)}_{-0.070(0.123)}$ & $-3.409^{+0.209(0.348)}_{-0.229(0.377)}$ & $0.337^{+0.032(0.061)}_{-0.040(0.058)}$ & $0.754^{+0.324(0.517)}_{-0.241(0.356)}$ & $-0.653^{+0.821(1.697)}_{-0.402(0.693)}$ & $-1$  & $0$\\
\hline
\multicolumn{7}{c}{$\omega_0$CDM} \\
\cline{1-7}
$1.476^{+0.085(0.137)}_{-0.081(0.145)}$ & $-3.411^{+0.182(0.376)}_{-0.242(0.376)}$ & $0.335^{+0.034(0.063)}_{-0.034(0.056)}$ & $0.763^{+0.226(0.433)}_{-0.189(0.340)}$ & $-$ & $-3.259^{+1.665(3.480)}_{-1.985(3.903)}$ &
$-$ \\
$1.458^{+0.082(0.144)}_{-0.089(0.148)}$ & $-3.390^{+0.214(0.360)}_{-0.219(0.373)}$ & $0.336^{+0.032(0.056)}_{-0.042(0.058)}$ & $0.850^{+0.340(0.534)}_{-0.296(0.539)}$ & $-0.549^{+0.515(1.286)}_{-0.440(0.634)}$ & $-2.082^{+2.098(2.534)}_{-1.359(4.835)}$ &
$-$ \\
\hline
\multicolumn{7}{c}{$\omega_0\omega_1$CDM}\\
\cline{1-7}
$1.486^{+0.073(0.125)}_{-0.101(0.155)}$ & $-3.433^{+0.221(0.358)}_{-0.220(0.345)}$ & $0.334^{+0.034(0.059)}_{-0.035(0.053)}$ & $0.728^{+0.257(0.505)}_{-0.175(0.371)}$ & $-$ & $-2.835^{+2.807(3.685)}_{-2.637(4.944)}$ &
$-0.377^{+5.419(6.492)}_{-2.783(3.544)}$ \\
$1.446^{+0.084(0.148)}_{-0.085(0.143)}$ & $-3.381^{+0.243(0.397)}_{-0.211(0.378)}$ & $0.330^{+0.040(0.062)}_{-0.036(0.052)}$ & $0.901^{+0.417(0.578)}_{-0.372(0.596)}$ & $-0.338^{+0.699(0.956)}_{-1.330(1.661)}$ & $-1.127^{+1.631(2.531)}_{-2.128(2.635)}$ &
$+3.672^{+2.216(2.566)}_{-3.267(4.032)}$ \\
\hline
\end{tabular}
\caption{Best-fit correlation and cosmological parameters at $1$-$\sigma$ ($2$-$\sigma$) got from the MCMC simulation of the $L_p-E_{p}$ correlation, calibrated with (upper part) and without (lower part) the DESI LRG1 data point.}
\label{tab:DESIBAOYonetoku1}
\end{table*}

We begin with the $\Lambda$CDM scenario.
\begin{enumerate}
\item[-]  When our MCMC computation considers the whole DESI data set, in the flat case our value of the matter density $\Omega_m$ agrees with both $\Omega_m^P=0.315\pm 0.007$ \cite{2021A&A...652C...4P} and $\Omega_m^S=0.299\pm0.016$ \cite{2021PhRvD.103h3533A} at $1$-$\sigma$. When the LRG1 data point is excluded, the deduced $\Omega_m$ is not compatible with neither Planck or the SDSS.
\item[-] In the non-flat scenario, the inferred value of $\Omega_m$ is compatible with both $\Omega_m^P = 0.348^{+0.013}_{-0.014}$ \cite{2021A&A...652C...4P} and $\Omega_m^S = 0.285^{+0.168}_{-0.170}$ \cite{2021PhRvD.103h3533A}, when considering the LRG1 data. Contrary, excluding it, our inferred $\Omega_m$ is consistent at $2$-$\sigma$ only with $\Omega_m^S$.
\item[-] With the whole DESI data set, we infer a $\Omega_k$ that is compatible with both $\Omega_k^P = -0.011^{+0.013}_{-0.012}$ \cite{2021A&A...652C...4P} and $\Omega_k^S = 0.079^{+0.083}_{-0.100}$ \cite{2021PhRvD.103h3533A} only at $2$-$\sigma$.
On the other hand, the exclusion of the LRG1 data, increases the errors on $\Omega_k$, leading to a compatibility at $1$-$\sigma$ level with both Planck and SDSS.
\end{enumerate}

For the $\omega_0$CDM model we obtain the results below.
\begin{enumerate}
\item[-] Within the flat scenario, both including or removing the LRG1 data point, the inferred matter density $\Omega_m$ are not compatible with $\Omega_m^P$ and $\Omega_m^S$. Further, when we compare the $\omega_0$ derived from our computations with the expectation of the concordance paradigm, \textit{i.e.}, $\omega_0=-1$, we find an agreement at $2$-$\sigma$ only when the LRG1 data is excluded from our analysis.
\item[-] On the other hand, in the non-flat scenario our $\Omega_m$ is compatible at $1$-$\sigma$ only with $\Omega_m^S$, when we consider the whole BAO sample. Then, excluding LRG1, $\Omega_m$ is found to be compatible only at $2$-$\sigma$ with both with both $\Omega_m^P$ and $\Omega_m^S$. In this case, our $\omega_0$ is compatible with $\omega_0=-1$ in both analyses.
\item[-] Finally, the curvature parameter $\Omega_k$ agrees with both $\Omega_k^P$ and $\Omega_k^S$, with or without removing the LRG1 data, only at $2$-$\sigma$.
\end{enumerate}

For the $\omega_0\omega_1$CDM case, we infer the following results.
\begin{enumerate}
\item[-] For the flat scenario we find that our $\Omega_m$ agrees only with $\Omega_m^P = 0.315\pm 0.007$ at $2$-$\sigma$, when the whole DESI catalog is considered. The inferred values of $\omega_0$ agree at $1$-$\sigma$ with the $\Lambda$CDM model ($\omega_0 = -1$), in all the analyses; the $\omega_1$, instead, agree at $1$-$\sigma$ with the $\Lambda$CDM model ($\omega_1 = 0$), only when removing the LRG1 data from the analysis.
\item[-] In the non-flat case, our $\Omega_m$ agrees at $2$-$\sigma$ with $\Omega_m^P = 0.348^{+0.013}_{-0.014}$, with and without the inclusion of LRG1. The compatibility with $\Omega_m^S = 0.285^{+0.168}_{-0.170}$ is at $1$-$\sigma$ when including LRG1, while it is only at $2$-$\sigma$ when excluding it. Regarding $\omega_0$ and $\omega_1$, our estimates are compatible with the $\Lambda$CDM predictions at $1$-$\sigma$, when considering the whole DESI sample. Excluding the LRG1 data, $\omega_0$ agrees with $-1$ at $1$-$\sigma$, while $\omega_1$ is compatible with $0$ only at $2$-$\sigma$.
\item[-] Finally, our values of $\Omega_k$ are always compatible with $\Omega_k^P = -0.011^{+0.013}_{-0.012}$ and $\Omega_k^S = 0.079^{+0.083}_{-0.100}$.
\end{enumerate}

For all the cases discussed above it is worth to stress out that the higher values of $\Omega_m$ found in our computations are in agreement with the tendecy of GRBs to boost the value of the matter density \cite{2021JCAP...09..042K}.

\subsection{Results for the $L_0-E_p-T$ correlation function}

\begin{table*}[t]
\footnotesize
\centering
\setlength{\tabcolsep}{0.25em}
\renewcommand{\arraystretch}{1.6}
\begin{tabular}{ccccccc}
\hline\hline
\multicolumn{7}{c}{$L_0-E_p-T$ correlation}\\
\hline\hline
$a$ & $b$ & $\sigma$ & $\Omega_m$ & $\Omega_k$ & $\omega_0$ & $\omega_1$\\
\hline
\multicolumn{7}{c}{With LRG1}\\
\hline
\multicolumn{7}{c}{$\Lambda$CDM} \\
\cline{1-7}
$0.823^{+0.073(0.121)}_{-0.087(0.141)}$ & $49.595^{+0.237(0.380)}_{-0.160(0.289)}$ & $0.390^{+0.030(0.053)}_{-0.029(0.046)}$ & $0.410^{+0.152(0.269)}_{-0.117(0.177)}$ & $-$ & $-1$ & $0$ \\
$0.785^{+0.101(0.158)}_{-0.068(0.119)}$ & $49.680^{+0.204(0.329)}_{-0.224(0.373)}$ & $0.388^{+0.030(0.048)}_{-0.028(0.045)}$ & $0.425^{+0.156(0.264)}_{-0.116(0.186)}$ & $-0.346^{+0.337(0.698)}_{-0.287(0.414)}$ & $-1$  & $0$\\
\hline
\multicolumn{7}{c}{$\omega_0$CDM} \\
\cline{1-7}
$0.796^{+0.084(0.143)}_{-0.068(0.118)}$ & $49.734^{+0.154(0.278)}_{-0.239(0.385)}$ & $0.386^{+0.029(0.048)}_{-0.031(0.047)}$ & $0.485^{+0.122(0.204)}_{-0.085(0.145)}$ & $-$ & $-6.405^{+4.243(5.296)}_{-1.418(\rm unc)}$ &
$-$ \\
$0.783^{+0.081(0.134)}_{-0.083(0.137)}$ & $49.727^{+0.231(0.374)}_{-0.190(0.334)}$ & $0.383^{+0.031(0.050)}_{-0.029(0.046)}$ & $0.611^{+0.151(0.275)}_{-0.162(0.264)}$ & $-0.324^{+0.260(0.494)}_{-0.268(0.414)}$ & $-3.611^{+2.026(3.167)}_{-3.238(3.894)}$ &
$-$ \\
\hline
\multicolumn{7}{c}{$\omega_0\omega_1$CDM}\\
\cline{1-7}
$0.807^{+0.076(0.130)}_{-0.086(0.129)}$ & $49.690^{+0.214(0.327)}_{-0.197(0.326)}$ & $0.384^{+0.028(0.048)}_{-0.030(0.046)}$ & $0.505^{+0.094(0.166)}_{-0.112(0.157)}$ & $-$ & $-4.950^{+3.099(3.828)}_{-1.746(2.806)}$ &
$-4.866^{+2.708(3.791)}_{-1.663(2.339)}$ \\
$0.768^{+0.086(0.149)}_{-0.077(0.129)}$ & $49.777^{+0.212(0.350)}_{-0.201(0.364)}$ & $0.384^{+0.028(0.050)}_{-0.032(0.047)}$ & $0.613^{+0.196(0.277)}_{-0.139(0.256)}$ & $-0.302^{+0.352(0.645)}_{-0.221(0.351)}$ & $-7.667^{+3.610(4.310)}_{\rm unc}$ &
$+2.169^{+2.824(3.539)}_{-2.447(3.018)}$ \\
\hline
\hline
\multicolumn{7}{c}{Without LRG1}\\
\hline
\multicolumn{7}{c}{$\Lambda$CDM} \\
\cline{1-7}
$0.810^{+0.082(0.132)}_{-0.085(0.144)}$ & $49.548^{+0.261(0.412)}_{-0.170(0.298)}$ & $0.406^{+0.032(0.052)}_{-0.028(0.044)}$ & $0.703^{+0.251(0.500)}_{-0.179(0.281)}$ & $-$ & $-1$ & $0$ \\
$0.782^{+0.095(0.145)}_{-0.073(0.128)}$ & $49.633^{+0.215(0.363)}_{-0.212(0.346)}$ & $0.407^{+0.030(0.049)}_{-0.030(0.043)}$ & $0.738^{+0.256(0.421)}_{-0.190(0.309)}$ & $-0.409^{+0.638(1.216)}_{-0.449(0.656)}$ & $-1$  & $0$\\
\hline
\multicolumn{7}{c}{$\omega_0$CDM} \\
\cline{1-7}
$0.806^{+0.083(0.144)}_{-0.086(0.143)}$ & $49.595^{+0.245(0.394)}_{-0.196(0.358)}$ & $0.410^{+0.029(0.048)}_{-0.033(0.047)}$ & $0.762^{+0.169(0.295)}_{-0.175(0.300)}$ & $-$ & $-4.172^{+3.239(4.288)}_{-3.727(4.876)}$ &
$-$ \\
$0.803^{+0.059(0.115)}_{-0.097(0.149)}$ & $49.655^{+0.216(0.348)}_{-0.191(0.321)}$ & $0.403^{+0.033(0.055)}_{-0.026(0.043)}$ & $0.831^{+0.291(0.478)}_{-0.200(0.304)}$ & $-0.362^{+0.503(0.880)}_{-0.448(0.710)}$ & $-3.004^{+1.425(2.461)}_{-2.113(2.869)}$ &
$-$ \\
\hline
\multicolumn{7}{c}{$\omega_0\omega_1$CDM}\\
\cline{1-7}
$0.812^{+0.073(0.127)}_{-0.081(0.144)}$ & $49.573^{+0.228(0.371)}_{-0.174(0.308)}$ & $0.406^{+0.032(0.050)}_{-0.029(0.045)}$ & $0.719^{+0.277(0.763)}_{-0.211(0.396)}$ & $-$ & $-2.036^{+2.392(4.266)}_{-3.530(\rm unc)}$ &
$+4.274^{+1.780(\rm unc)}_{-6.040(11.064)}$ \\
$0.803^{+0.068(0.128)}_{-0.086(0.144)}$ & $49.617^{+0.211(0.358)}_{-0.184(0.334)}$ & $0.408^{+0.028(0.047)}_{-0.032(0.049)}$ & $0.592^{+0.172(0.451)}_{-0.068(0.230)}$ & $-0.790^{+2.598(3.871)}_{-1.756(2.210)}$ & $-1.129^{+0.821(1.135)}_{-3.006(3.711)}$ &
$-1.417^{+0.661(0.944)}_{-0.792(1.082)}$ \\
\hline
\end{tabular}
\caption{Best-fit correlation and cosmological parameters at $1$-$\sigma$ ($2$-$\sigma$) from the MCMC simulation of $L_0-E_{p}-T$ correlation, calibrated with (upper part) and without (lower part) the DESI LRG1 data point.}
\label{tab:DESIBAOCombo1}
\end{table*}

The best-fit correlation and cosmological parameters inferred by considering the calibrated $L_0-E_p-T$ correlation are reported in Tab. \ref{tab:DESIBAOCombo1}, while in Figs. \ref{plot:combofull}-\ref{plot:combonolrg1} the contour plots are displayed.
Again, results are  differentiated between flat and non-flat cosmological models and whether the LRG1 data point is included or not.

For the $\Lambda$CDM model, we list the conclusions below.
\begin{enumerate}
\item[-] In the flat scenario, the value of $\Omega_m$ inferred from the whole DESI sample is compatible with both $\Omega_m^P = 0.315\pm 0.007$ and $\Omega_m^S = 0.299\pm 0.016$ at $1$-$\sigma$. Contrary, when we exclude the LRG1 data point, our $\Omega_m$ is incompatible with both the matter densities from Planck or the SDSS.
\item[-] In the non-flat case, our $\Omega_m$ is again compatible with  $\Omega_m^P=0.348^{+0.013}_{-0.014}$ and $\Omega_m^S=0.285^{+0.168}_{-0.170}$ when the whole DESI sample is accounted for. On the other hand, excluding the LRG1 point, this time we have an agreement at $2$-$\sigma$ with $\Omega_m^S$ only.
\item[-] The inferred curvature parameter $\Omega_k$ is always compatible with both  $\Omega_k^P$ and $\Omega_k^S$.
\end{enumerate}

For the $\omega_0$CDM case, we found the following results.
\begin{enumerate}
\item[-] In the flat case, in all the analyses, our $\Omega_m$ is incompatible with either $\Omega_m^P$ or $\Omega_m^S$. The barotropic factor $\omega_0$, on the other hand, is slightly compatible with $\omega_0=-1$ at $2$-$\sigma$, when the complete data set is considered. The compatibility is at $1$-$\sigma$, when the LRG1 data is excluded.
\item[-] Including the curvature, with the whole DESI sample, our $\Omega_m$ appears to be compatible at $2$-$\sigma$ with both the values inferred from Planck and the SDSS. Excluding the LRG1 data, we do not have compatibility with $\Omega_m^P$ and $\Omega_m^S$. The barotropic factor $\omega_0$ is in agreement at $2$-$\sigma$ with $\omega_0 = -1$, both including or not the LRG1 data.
\item[-] Finally, the curvature $\Omega_k$ agrees at $2$-$\sigma$ with both $\Omega_k^P$ and $\Omega_k^S$, when the LRG1 is included, and at $1$-$\sigma$, when it is excluded.
 \end{enumerate}

We conclude our analysis with the $\omega_0\omega_1$CDM scenario.
\begin{enumerate}
\item[-] In the flat scenario, our $\Omega_m$ is not compatible with either $\Omega_m^P$ and $\Omega_m^S$ in both analyses, with and without LRG1. Our values of $\omega_0$ and $\omega_1$ are compatible with the concordance paradigm expectations at $1$-$\sigma$ only when excluding the LRG1 data point. When including it, only $\omega_0$ is slightly in agreement at $2$-$\sigma$ with $\omega_0 = -1$, while $\omega_1$ does not agree with $\omega_1 = 0$.
\item[-] In the non-flat case, our $\Omega_m$ agrees at $2$-$\sigma$ with $\Omega_m^S$ in both analyses, with and without LRG1. For $\omega_0$, the agreement with $\omega_0 = -1$ at $1$-$\sigma$ occurs only  when the LRG1-free sample is considered. For $\omega_1$ we have compatibility with $\omega_1 = 0$ only when we include the LRG1 data point.
\item[-] The curvature parameter $\Omega_k$ appears in agreement at $1$-$\sigma$ with both $\Omega_k^P$ and $\Omega_k^S$, in all the analyses.
\end{enumerate}

As for the $L_p-E_p$ correlation, we stress that the higher values inferred for $\Omega_m$ are in line with the trend of GRBs of boosting the value of the matter density when they are not combined with other probes \cite{2021JCAP...09..042K}.

\subsection{Criteria of model selection}

To address the goodness of our numerical analyses, we here compute model selection criteria aiming at understanding if dynamical dark energy may be favored, comparing our results with those of DESI collaboration.

Hence, we hereafter use the Akaike information criterion (AIC), the corrected AIC (AICc), the Bayesian information criterium (BIC) and the deviance information criterion (DIC) \cite{Akaike:1998zah, akaike2011akaike, 2007MNRAS.377L..74L, 2006PhRvD..74b3503K, sugiura1978further}, respectively,
\begin{subequations}
\label{modelcrit}
    \begin{align}
        &\text{AIC}\equiv -2\ln\mathcal{L}_m+2d,&\\
        &\text{AICc}\equiv \frac{2d(d+1)}{N-d-1}+\text{AIC},&\\
        &\text{BIC}\equiv -2\ln\mathcal{L}_m+2d\ln(N),&\\
        &\text{DIC}\equiv 2\ln\mathcal{L}_m+2\langle-2\ln\mathcal{L}\rangle,&
    \end{align}
\end{subequations}
where $\ln\mathcal{L}_m$ is the maximum log-likelihood, $d$ represents the free parameters of each cosmological model, $N$ is the number of data points that follows either the $L_p-E_p$ or the $L_0-E_p-T$ correlations and $\langle-2\ln\mathcal{L}\rangle$ is the average taken over the posterior distribution.

To argue which model appears statistically favored from our analyses, we evaluate the differences between each value inferred from Eqs. \eqref{modelcrit} and the lowest ones, labelled as $Y_0$ for each model,
%
\begin{equation}
    \Delta Y = Y_i-Y_0, \quad  Y_i = \text{\{AIC, AICc, BIC, DIC\}}.
\end{equation}

It can be seen from Tab. \ref{tab:modelsel} that neither the $L_p-E_p$ or the $L_0-E_p-T$ correlation functions seem to show a preference towards one cosmological model in particular, albeit the $\Lambda$CDM model seems to be the least favored model, when the LRG1 data point is included.

 \begin{table*}
\centering
\setlength{\tabcolsep}{.8em}
\renewcommand{\arraystretch}{1.2}
\begin{tabular}{l|lcccccccccc}
\hline
\multicolumn{11}{c}{$L_p-E_p$}\\
\hline
   & & $-\ln \mathcal L_m$ & \text{AIC} & AICc & BIC & DIC & $\Delta$AIC & $\Delta$AICc & $\Delta$BIC & $\Delta$DIC\\
\hline
\multirow{6}{*}{\rm With\ LRG1} &
$\Lambda$CDM ($\Omega_k=0$) & $162.17$ & $332$ & $333$ & $343$ & $333$ & $6$ & $5$ & $1$ & $5$ \\
&
$\Lambda$CDM ($\Omega_k\neq0$) & $159.47$ & $329$ & $329$ & $342$ & $330$ & $2$ & $2$ & $1$ & $3$ \\
&
$\omega_0$CDM ($\Omega_k=0$) & $159.08$ & $328$ & $329$ & $341$ & $327$ & $1$ & $1$ & $0$ & $0$\\
&
$\omega_0$CDM ($\Omega_k\neq0$) & $157.33$ & $327$ & $327$ & $342$ & $328$ & $0$ & $0$ & $1$ & $0.5$ \\
&
$\omega_0\omega_1$CDM ($\Omega_k=0$) &   $159.00$ & $330$ & $331$ & $346$ & $329$ & $3$ & $3$ & $4$ & $1$\\
&
$\omega_0\omega_1$CDM ($\Omega_k\neq0$) &  $157.23$ & $328$ & $330$ & $347$ & $328$ & $2$ & $2$ & $5$ & $1$ \\
\hline
\multirow{6}{*}{\rm Without\ LRG1} &
$\Lambda$CDM ($\Omega_k=0$) & $160.41$ & $329$ & $329$ & $339$ & $329$ & $1$ & $1$ & $0$ & $1$ \\
&
$\Lambda$CDM ($\Omega_k\neq0$) & $158.92$ & $328$ & $328$ & $341$ & $330$ & $0.3$ & $0$ & $2$ & $1$ \\
&
$\omega_0$CDM ($\Omega_k=0$) & $159.59$  & $329$ & $330$ & $342$ & $328$ & $2$ & $1$ & $3$ & $0$\\
&
$\omega_0$CDM ($\Omega_k\neq0$) & $158.18$ & $328$ & $329$ & $344$ & $329$ & $1$ & $1$ & $5$ & $1$ \\
&
$\omega_0\omega_1$CDM ($\Omega_k=0$) &  $159.09$ & $330$ & $331$ & $346$ & $330$ & $3$ & $3$ & $7$ & $1$\\
&
$\omega_0\omega_1$CDM ($\Omega_k\neq0$) & $156.74$ & $327$ & $329$ & $346$ & $330$ & $0$ & $0.2$ & $6$ & $2$ \\
\hline
\multicolumn{11}{c}{$L_0-E_p-T$}\\
\hline
\multirow{6}{*}{\rm With\ LRG1} &
$\Lambda$CDM ($\Omega_k=0$) & $345.63$ & $699$ & $699$ & $712$ & $700$ & $7$ & $7$ & $3$ & $8$ \\
&
$\Lambda$CDM ($\Omega_k\neq0$) & $344.19$ & $698$ & $699$ & $714$ & $699$ & $6$ & $6$ & $6$ & $8$ \\
&
$\omega_0$CDM ($\Omega_k=0$) & $341.25$ & $692$ & $693$ & $708$ & $694$ & $0.3$ & $0.1$ & $0$ & $2$\\
&
$\omega_0$CDM ($\Omega_k\neq0$) & $340.11$ & $692$ & $693$ & $711$ & $694$ & $0$ & $0$ & $3$ & $2$ \\
&
$\omega_0\omega_1$CDM ($\Omega_k=0$) &   $341.36$ & $695$ & $695$ & $714$ & $694$ & $2$ & $2$ & $5$ & $2$\\
&
$\omega_0\omega_1$CDM ($\Omega_k\neq0$) &  $339.42$ & $693$ & $693$ & $715$ & $691$ & $1$ & $1$ & $7$ & $0$ \\
\hline
\multirow{6}{*}{\rm Without\ LRG1} &
$\Lambda$CDM ($\Omega_k=0$) & $355.52$ & $719$ & $719$ & $732$ & $719$ & $3$ & $2$ & $0$ & $1$ \\
&
$\Lambda$CDM ($\Omega_k\neq0$) & $354.94$ & $720$ & $720$ & $736$ & $720$ & $4$ & $3$ & $4$ & $1$ \\
&
$\omega_0$CDM ($\Omega_k=0$) & $354.50$  & $719$ & $719$ & $735$ & $719$ & $3$ & $3$ & $3$ & $0$\\
&
$\omega_0$CDM ($\Omega_k\neq0$) & $353.64$ & $719$ & $720$ & $738$ & $719$ & $3$ & $3$ & $7$ & $0.4$ \\
&
$\omega_0\omega_1$CDM ($\Omega_k=0$) &  $353.85$ & $720$ & $720$ & $739$ & $721$ & $4$ & $3$ & $7$ & $2$\\
&
$\omega_0\omega_1$CDM ($\Omega_k\neq0$) & $351.02$ & $716$ & $717$ & $738$ & $721$ & $0$ & $0$ & $7$ & $2$ \\
\hline
\end{tabular}
\caption{Comparison, based on the selection criteria, between flat and non-flat $\Lambda$CDM, $\omega_0$CDM and $\omega_0\omega_1$CDM models. The upper panel shows the criteria for the $L_p-E_p$ correlation, while the lower panel shows those for the $L_0-E_p-T$ correlation.}
\label{tab:modelsel}
\end{table*}

\section{Conclusion}\label{sec5}

The early release prompted by the DESI Collaboration seems to point to a {\it  renaissance} of dynamical dark energy, since the first analysis leans in the direction of a tension with the current cosmological scenario, where the dark energy fluid guiding the late accelerated expansion acts as a cosmological constant.

Since then, many works used the DESI-BAO data points, alone or in combination with other probes, to back up or dismiss this claim \cite{2024arXiv240408633C, 2024arXiv240412068C, 2024A&A...690A..40L, 2024arXiv240500502P, 2024arXiv240504216C, 2024arXiv240513588L, 2024JCAP...09..062D, 2024arXiv240519178M, 2024arXiv240706586P, 2024arXiv240802365J, 2024arXiv240802536A, 2024arXiv240415232G}. Some of them even pointed out to data points that could be problematic and thus affecting the overall analysis \cite{2024arXiv240408633C, 2024arXiv240415232G, 2024arXiv240502168W, 2024arXiv240610202L, 2024arXiv240702558C}. In this work we consider this claim regarding the LRG1  data point at $z_{eff} =0.51$ \cite{2024arXiv240408633C}.

In our analyses, we utilize two GRB correlations, {\it i.e.}, the $L_p-E_p$ and $L_0-E_p-T$, since they seem to point to a slightly evolving dark energy rather than a genuine cosmological constant \cite{2020A&A...641A.174L,2021MNRAS.503.4581L}. Our recipe consists firstly in calibrating the correlations through OHD and DESI-BAO data sets, making use of {\it B\`ezier} polynomials to avoid the {\it circularity problem} and turn them into distance indicators.

Afterwards, we involve the two correlations into two independent MCMC analyses, based on the Metropolis-Hastings algorithm \cite{1953JChPh..21.1087M, 1970Bimka..57...97H}. In the first analysis we consider the whole DESI data set, while in the second we exclude the LRG1 data point. All of this was done to analyze the three dark energy scenario taken under study by the DESI Collaboration, specifically:
\begin{itemize}
\item [-] a cosmological constant $\Lambda$, as purported by the concordance paradigm;
\item [-] a scalar field labeled as quintessence with a constant barotropic factor $\omega_0$;
\item [-] a dynamical dark energy, the scenario favored by DESI, which assumes an evolving barotropic factor.
\end{itemize}

Here we summarize the outcomes of our analyses for both the  $L_p-E_p$ and $L_0-E_p-T$ correlations.

Discussing the $L_p-E_p$ correlation first, for the $\Lambda$CDM model the value of $\Omega_m$ inferred from our computations agree with both matter densities from Planck $\Omega_m^P$ and from the SDSS $\Omega_m^S$ in both the flat and non-flat scenario when we consider the whole DESI data set. The curvature parameter $\Omega_k$ shows an agreement with the values inferred from Planck and the SDSS, \textit{i.e.}, $\Omega_k^P = -0.011^{+0.013}_{-0.012}$ and $\Omega_k^S = 0.079^{+0.083}_{-0.100}$ only when the whole DESI data set is considered.

For the flat $\omega_0$CDM scenario, on the other hand, the compatibility with our $\Omega_m$ is not reached in both the cases taken under study while the barotropic factor $\omega_0$ agrees only at $2$-$\sigma$ with the expectation of the $\Lambda$CDM model, \textit{i.e.}, $\omega_0=-1$ when the LRG1 data is excluded.
In the non-flat case, our $\Omega_m$ is compatible at $2$-$\sigma$ with both $\Omega_m^S = 0.285^{+0.168}_{-0.170}$ and $\Omega_m^P = 0.348^{+0.013}_{-0.014}$, when the LRG1 is excluded.
Excluding the LRG1 data, we do not have compatibility with $\Omega_m^P$ and $\Omega_m^S$. The barotropic factor $\omega_0$ is compatible at $2$-$\sigma$ with $\omega_0 = -1$, both including or not the LRG1 data.
The curvature parameter $\Omega_k$ agrees at $2$-$\sigma$ with both $\Omega_k^P$ and $\Omega_k^S$, when the LRG1 is included, and at $1$-$\sigma$, when it is excluded.

In the $\omega_0\omega_1$CDM scenario we found that only $\Omega_m^P$ is compatible at both $2$-$\sigma$ with ours, when the LRG1 data is considered. Also, the parameters $\omega_0$ and $\omega_1$ are in agreement with $\omega_0=-1$ and $\omega_1=0$ from the $\Lambda$CDM scenario when the LRG1 data is excluded. When $\Omega_k\neq 0$, our matter density is in agreement at $2$-$\sigma$ with both Planck and the SDSS values, when we consider all DESI data. This compatibility reaches also $1$-$\sigma$ with $\Omega_m^S$ when the LRG1 point is included in our analysis. In this case, $\omega_0$ and $\omega_1$ agrees with the concordance paradigm only with the inclusion of LRG1. The curvature, on the other hand, is compatible with both $\Omega_k^P$ and $\Omega_k^S$.

Switching to the $L_0-E_p-T$ correlation we found that for the flat (non-flat) $\Lambda$CDM model our $\Omega_m$ agrees with both $\Omega_m^P$ and $\Omega_m^S$ at $1$-$\sigma$ ($2$-$\sigma$) only when we consider the whole DESI data set. The curvature parameter $\Omega_k$ deduced from our computations agree with both values inferred from Planck and the SDSS.

For the flat $\omega_0$CDM case, our $\Omega_m$ does not agrees with $\Omega_m^P$ and $\Omega_m^S$, both including or excluding the LRG1 data point. The barotropic factor $\omega_0$ is compatible at $1$-$\sigma$ without the LRG1 data point. When curvature is considered, our $\Omega_m$ is in agreement at $2$-$\sigma$ with Planck and the SDSS values only when LRG1 is considered; $\omega_0$ agrees at $2$-$\sigma$ in both cases with the concordance paradigm expectation of $\omega_0 = -1$. The curvature parameter agrees at $2$-$\sigma$ only when we consider the whole DESI sample.

Finally, for the $\omega_0\omega_1$CDM scenario, we find incompatible values of $\Omega_m$, within the flat scenario, either considering or not the LRG1 data point. The values of $\omega_0$ and $\omega_1$ are compatible with the concordance paradigm when excluding the LRG1 data. In the non-flat case, our $\Omega_m$ agrees at $2$-$\sigma$ with $\Omega_m^S$, in both computations. The barotropic factor $\omega_0$ is compatible with $\omega_0 = -1$ at $1$-$\sigma$ only when excluding the LRG1 data point. Finally, the inferred curvature parameter $\Omega_k$ agrees with both values from Planck and the SDSS, in all the analyses.

Also, contrary to what we found in Ref. \cite{2024arXiv240802536A} it seems that the data point at $z_{eff} = 0.51$ does not dictate the behaviour of the matter density $\Omega_m$ considering that when using both the $L_p-E_p$ and $L_0-E_p-T$ correlations higher values of $\Omega_m$ are also found when the LRG1 data point is included. Moreover, if the $E_p-E_{iso}$ correlation showed a preference towards the flat $\Lambda$CDM model, in both our MCMC computations the considered correlations seem not to prefer a particular cosmological model, albeit the $\Lambda$CDM model seems to be the least favored one, when the LRG1 data point is included.
These last statements will be object of studies also in view of further data releases by the DESI Collaboration.

\acknowledgments

ACA acknowledges Istituto Nazionale di Fisica Nucleare (INFN) Sez. di Napoli, Iniziativa Specifica QGSKY.

\appendix

\begin{widetext}

\section{Contour plots}

Figs.~\ref{plot:yonetokufull}--\ref{plot:combonolrg1} show the contour plots obtained from our MCMC analyses, based on calibrated GRB correlations, for flat and non-flat $\Lambda$CDM, $\omega_0$CDM and $\omega_0\omega_1$CDM cosmological models.

\newpage

\begin{figure*}
\includegraphics[width=0.5\linewidth]{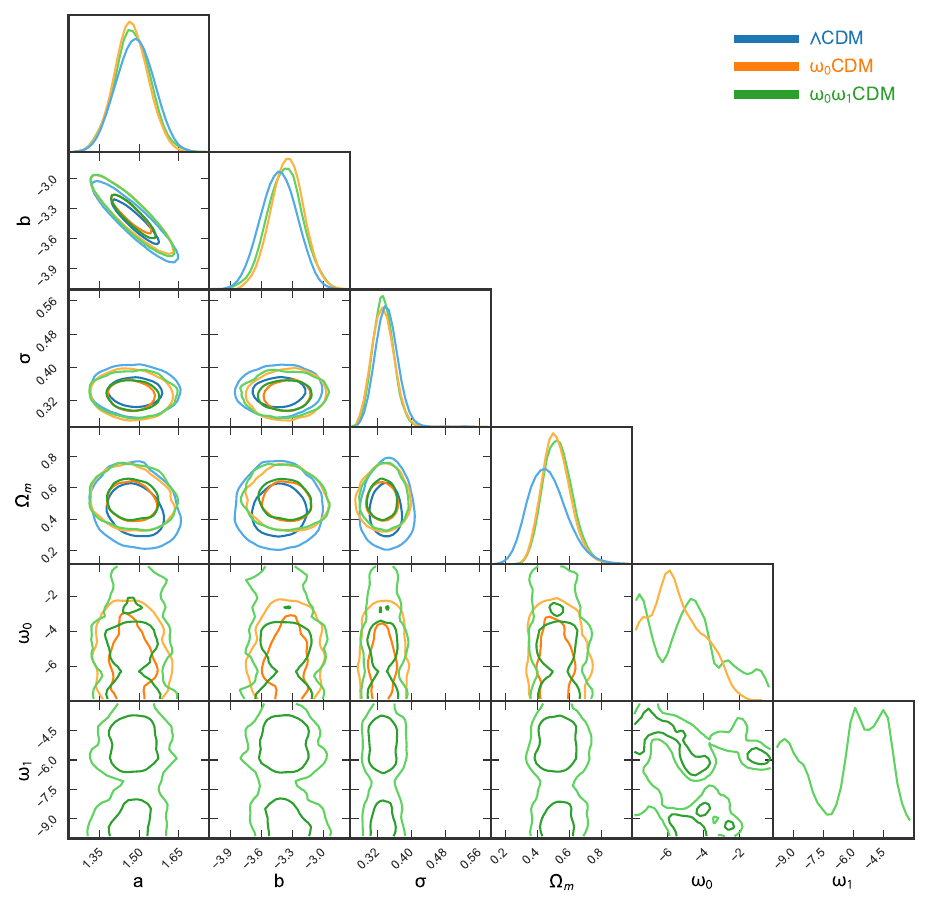}
\includegraphics[width=0.49\linewidth]{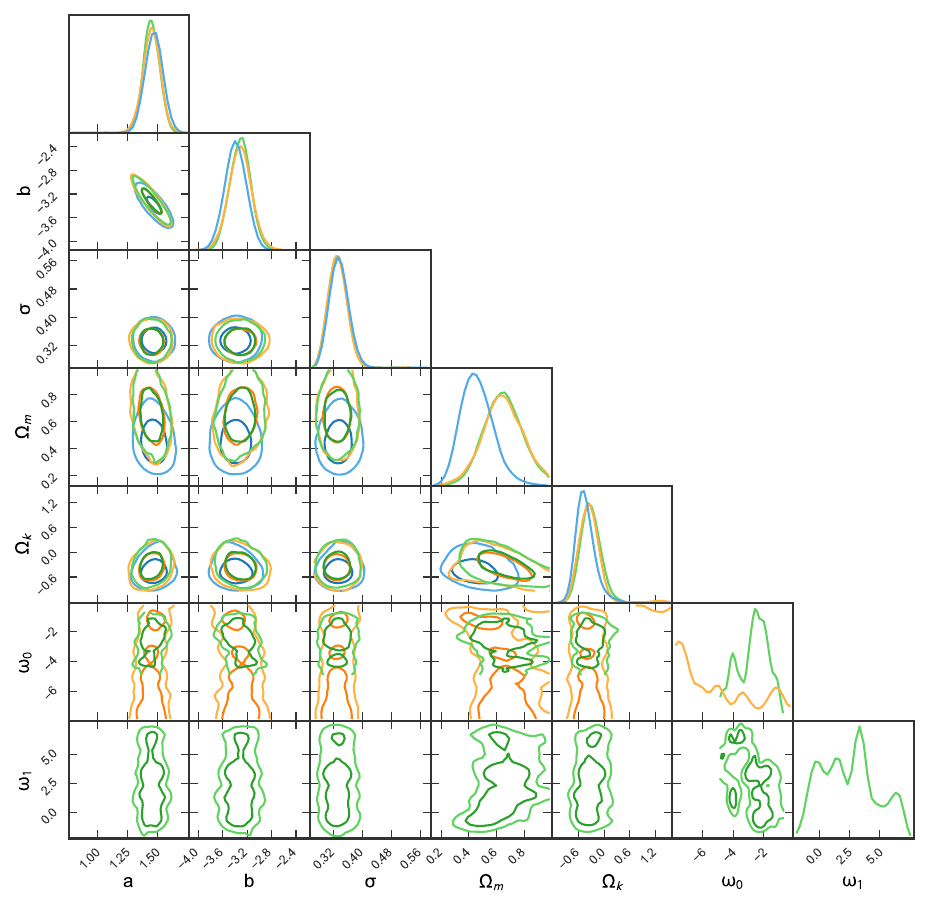}
\caption{Contour plots for the $L_p-E_p$ correlation functions when the all DESI sample is considered. The left panel shows the contours for the flat case while the right panel shows the contours for the non-flat case.}
\label{plot:yonetokufull}
\end{figure*}

\begin{figure*}
    \centering
    \includegraphics[width=0.5\linewidth]{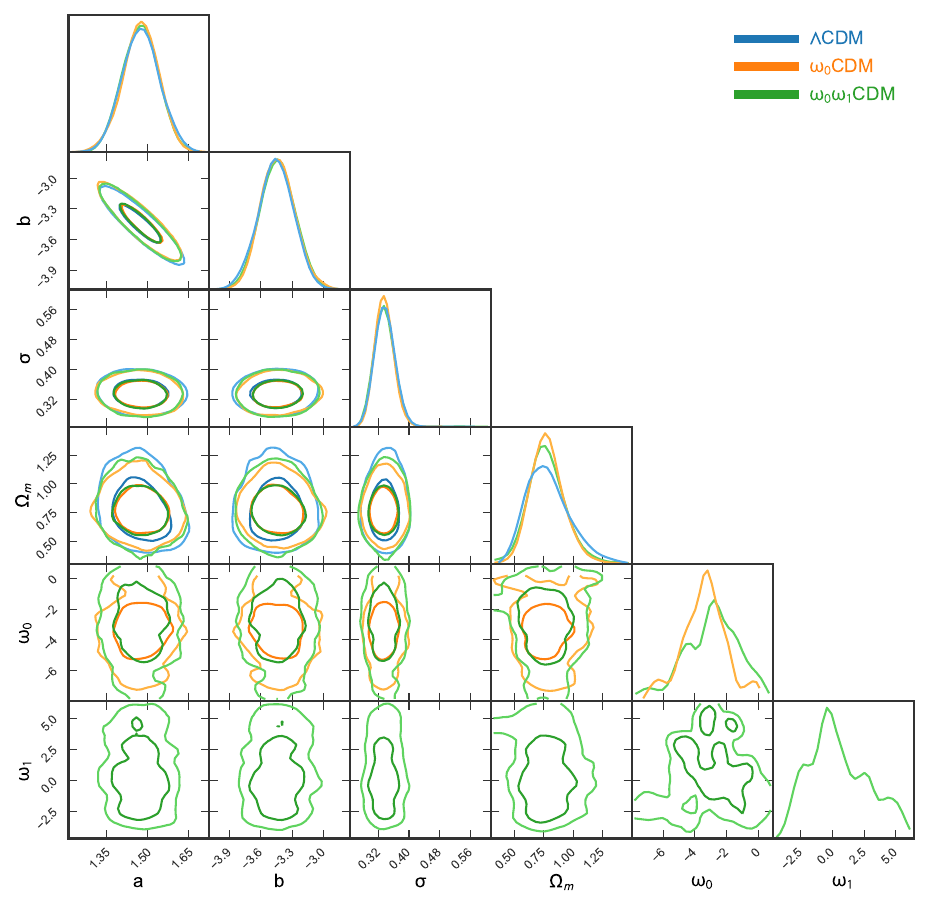}
    \includegraphics[width=0.49\linewidth]{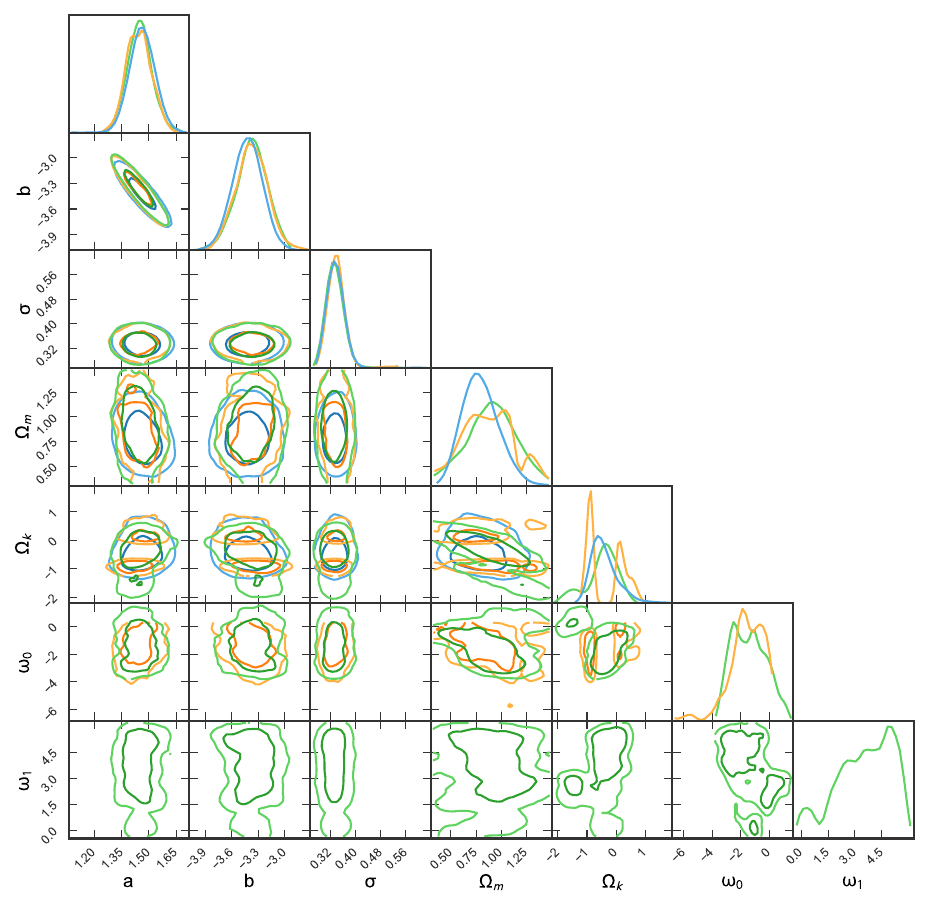}
\caption{Contour plots for the $L_p-E_p$ correlation functions when the LRG1 data point is excluded. The left panel shows the contours for the flat case while the right panel shows the contours for the non-flat case.}
\label{plot:yonetokunolrg1}
\end{figure*}

\begin{figure*}
\centering
\includegraphics[width=0.5\linewidth]{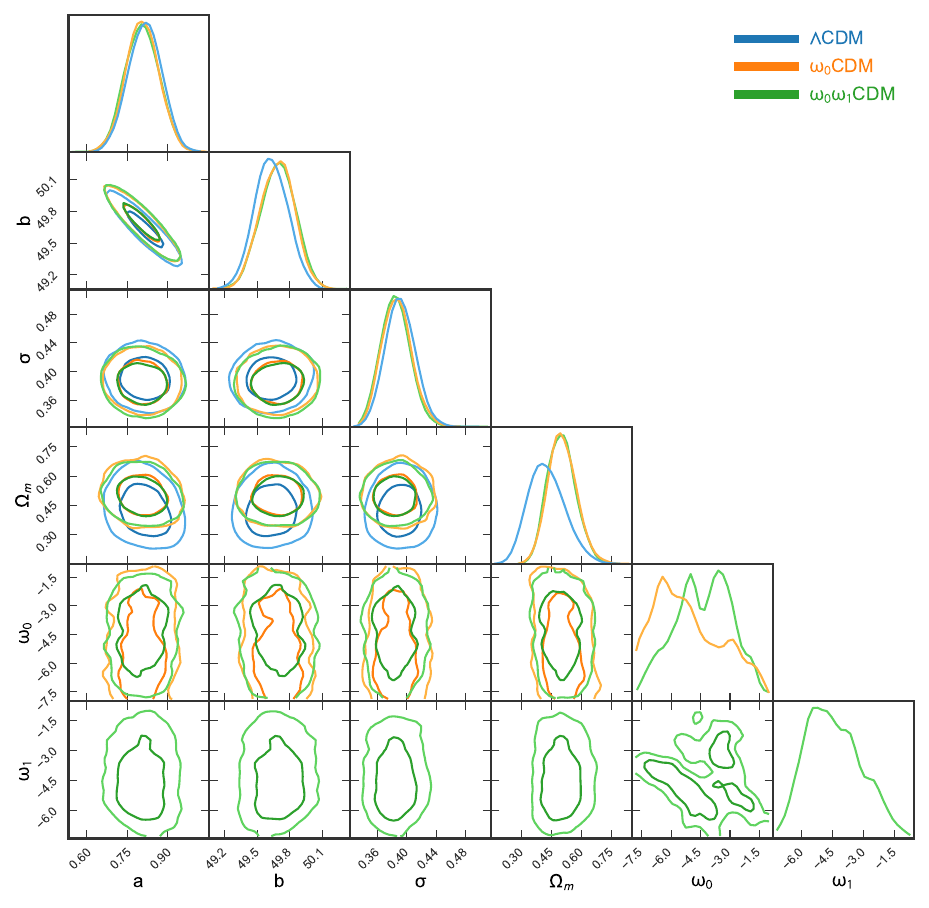}
\includegraphics[width=0.49\linewidth]{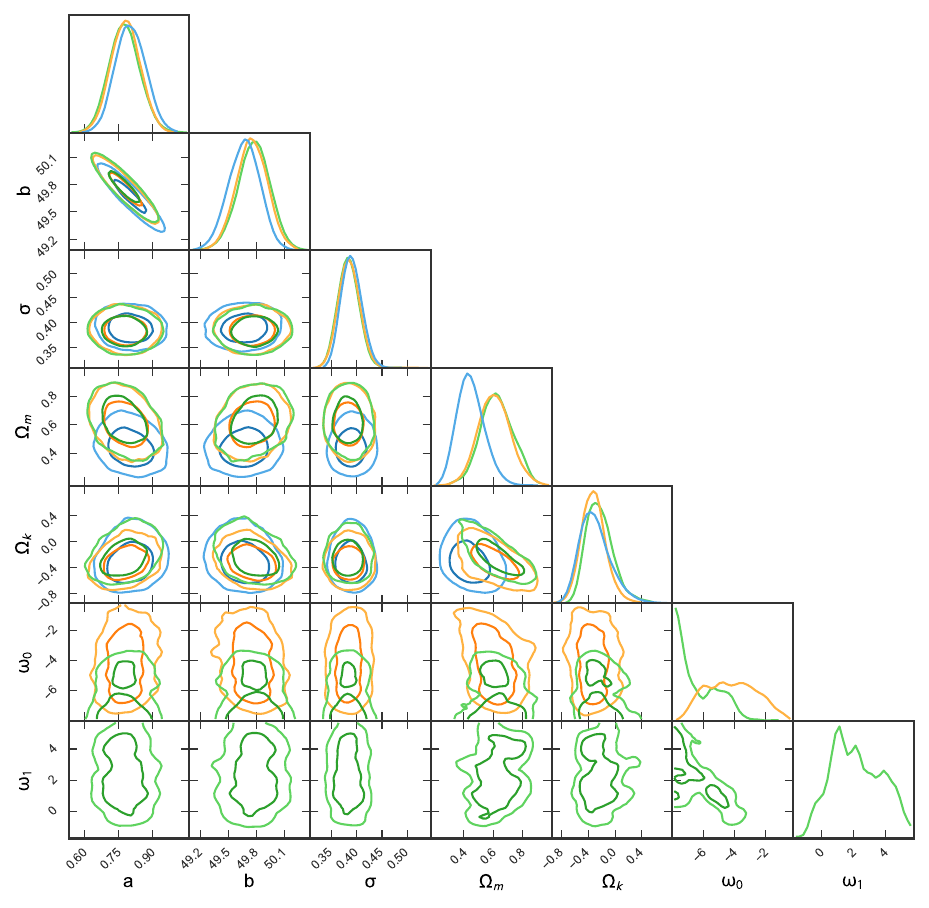}
\caption{Contour plots for the $L_0-E_p-T$ correlation function when the all DESI sample is considered. The left panel shows the contours for the flat case while the right panel shows the contours for the non-flat case.}
\label{plot:combofull}
\end{figure*}

\begin{figure*}
\centering
\includegraphics[width=0.5\linewidth]{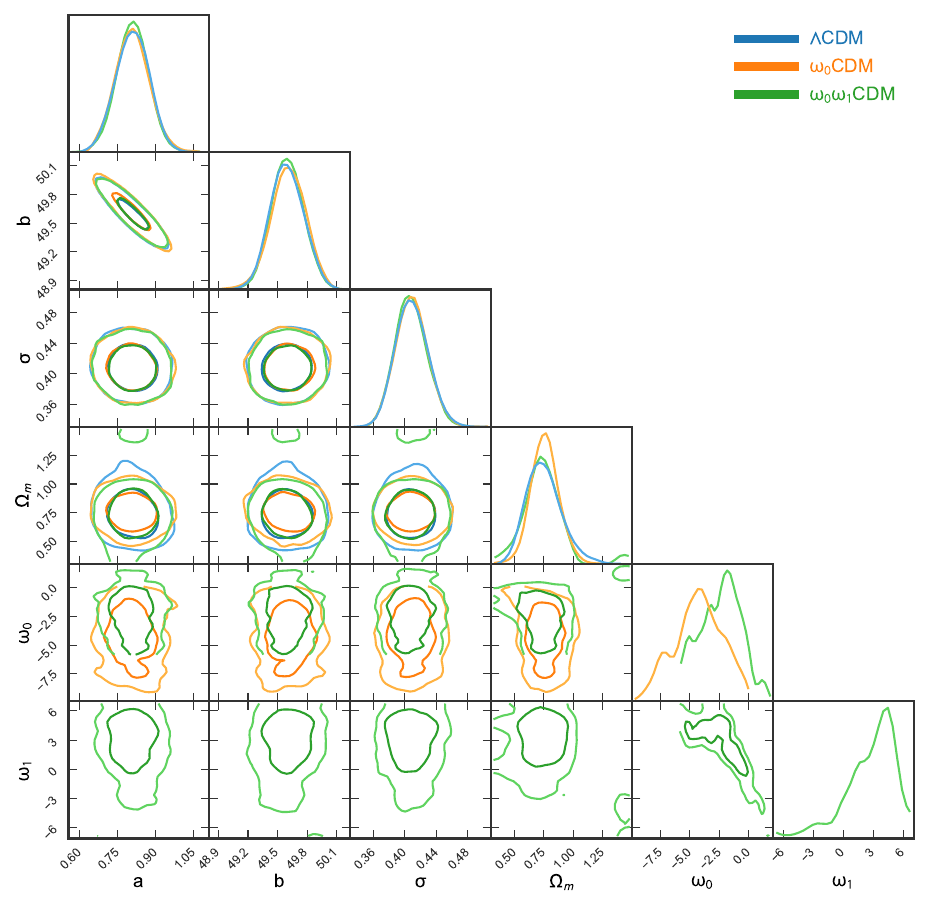}
\includegraphics[width=0.49\linewidth]{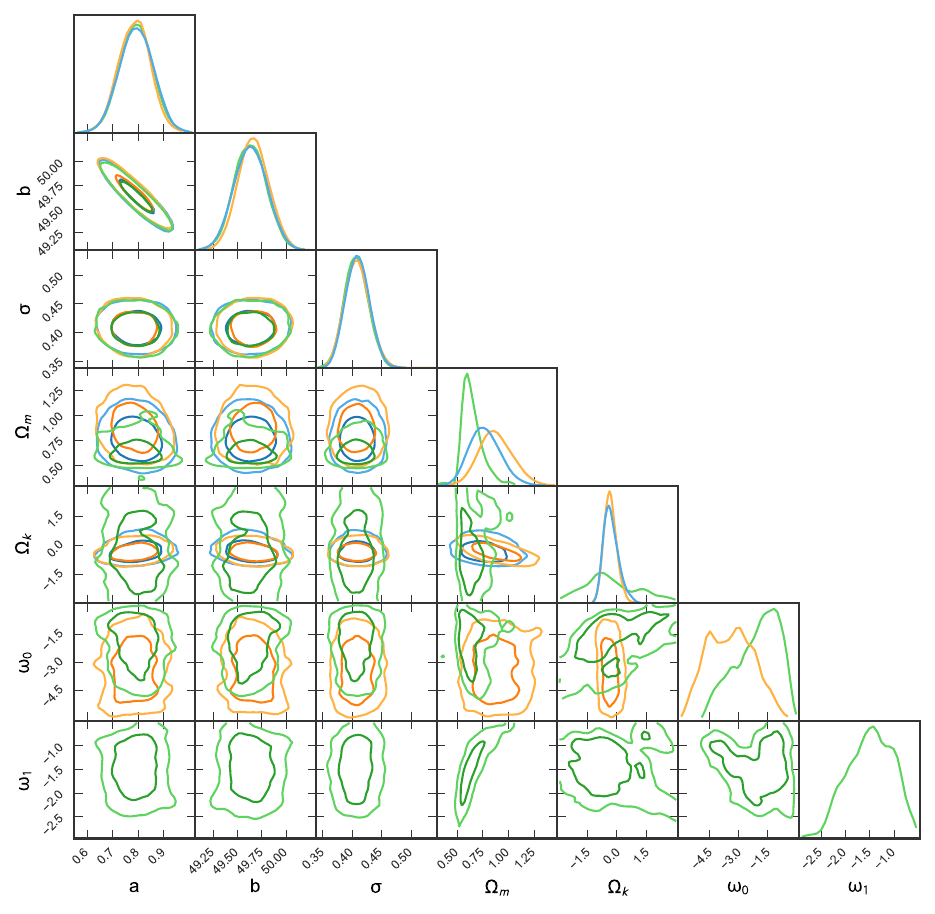}
\caption{Contour plots for the $L_0-E_p-T$ correlation function when the LRG1 data point is excluded. The right panel shows the contours for the flat case while the left panel shows the contours for the non-flat case.}
\label{plot:combonolrg1}
\end{figure*}

\end{widetext}

\end{document}